\documentclass[fleqn,10pt]{wlscirep}
\title{Fermi-Surface Topological Phase Transition and Horizontal Order-Parameter Nodes in CaFe$_2$As$_2$ Under Pressure}
\author[1,*]{R. S. Gonnelli}
\author[1]{D. Daghero}
\author[1]{M. Tortello}
\author[1]{G. A. Ummarino}
\author[2]{Z. Bukowski}
\author[3]{J. Karpinski}
\author[4]{P. G. Reuvekamp}
\author[4]{R. K. Kremer}
\author[5]{G. Profeta}
\author[6]{K. Suzuki}
\author[6]{K. Kuroki}

\affil[1]{Dipartimento di Scienza Applicata e Tecnologia, Politecnico di Torino, 10129 Torino, Italy}
\affil[2]{Institute of Low Temperature and Structure Research, Polish Academy of Sciences, 50-950 Wroc{\l}aw, Poland}
\affil[3]{Laboratory for Solid State Physics, Swiss Federal Institute of Technology (ETH), CH-8093 Z\"{u}rich, Switzerland}
\affil[4]{Max Planck Institute for Solid State Research, D-70569 Stuttgart, Germany}
\affil[5]{Dipartimento di Scienze Fisiche e Chimiche, Universit\`{a} dell'Aquila, 67100 Coppito (AQ) Italy}
\affil[6] {Department of Physics, Osaka University, Toyonaka, Osaka 560-0043, Japan}
\affil[*]{renato.gonnelli@polito.it}

\begin{abstract}
Iron-based compounds (IBS) display a surprising variety of superconducting properties that seems to arise from the strong sensitivity of these systems to tiny details of the lattice structure.  In this respect, systems that become superconducting under pressure, like  CaFe$_2$As$_2$, are of particular interest. Here we report on the first directional point-contact Andreev-reflection spectroscopy (PCARS) measurements on CaFe$_2$As$_2$ crystals under quasi-hydrostatic pressure, and on the interpretation of the results using a 3D model for Andreev reflection combined with ab-initio calculations of the Fermi surface (within the density functional theory) and of the order parameter symmetry (within a random-phase-approximation approach in a ten-orbital model). The almost perfect agreement between PCARS results at different pressures and theoretical predictions highlights the intimate connection between the changes in the lattice structure, a topological transition in the hole-like Fermi surface sheet, and the emergence on the same  sheet of an order parameter with a horizontal node line.
\end{abstract}

\begin{document}
\flushbottom
\maketitle
\thispagestyle{empty}

\section*{Introduction}
The picture of a superconducting pairing mediated by spin fluctuations (SF) in iron-based compounds (IBS) is justified by the proximity of superconductivity to an antiferromagnetic (AFM) state \cite{Lumsden} characterized by a $(\pi, \pi)$ ordering wavevector that connects holelike and electronlike sheets of the Fermi surface (FS). This mechanism, which would lead to the so-called $s\pm$  symmetry of the OP \cite{Mazin}, has been questioned by the discovery of superconducting compounds with only electronlike or holelike FS, such as  FeSe monolayers on SrTiO$_3$ \cite{Liu},  KFe$_2$As$_2$ \cite{Tafti,Li}, K$_{0.8}$Fe$_{1.7}$Se$_2$ \cite{Qian}.

The disappearance of a FS sheet (Lifshitz transition) was also found in CaFe$_2$As$_2$. When cooled at ambient pressure, CaFe$_2$As$_2$ shows a tetragonal (T) to orthorhombic (OR) structural transition, accompanied by the paramagnetic (PM) to AFM transition. Under moderate pressure, a tetragonal (T)  to collapsed tetragonal (cT) transition occurs on cooling \cite{Torikachvili,Park}. In the cT phase, stable at low temperature under hydrostatic pressure, the out-of-plane $c$ lattice constant is shorter by 10\% than in the T phase; angle-resolved photoemission spectroscopy (ARPES) \cite{Dhaka}, transport and Seebeck measurements \cite{Gofryk} as well as Density Functional Theory (DFT) calculations \cite{Tompsett} concur to indicate the absence of holelike FS around $\Gamma$. Inelastic neutron scattering experiments show no spin fluctuations \cite{Soh,Pratt}, and indeed this phase is not superconducting \cite{Yu}. However, superconductivity was observed in CaFe$_2$As$_2$ under pressure \cite{Torikachvili,Park} and ascribed to a paramagnetic ``non-collapsed'' tetragonal phase close to a magnetic instability \cite{Sanna}. Such a tetragonal phase was indeed observed experimentally and stabilized by non-hydrostatic pressure \cite{Prokes}.
Pressure can thus drive CaFe$_2$As$_2$ across a structural and electronic transition that makes it bridge the gap between ``usual'' multiband IBS and ``anomalous'' single-band ones, and is associated with the emergence of a superconducting phase with a relatively high $T_c$.

%To shed light on the role of the closeness of the cT phase to the superconducting phase, on the importance of the Lifshitz transition in the appearance of superconductivity in IBS, and on the effects of this transition on the symmetry of the OP, we designed a dedicated experimental study of the superconducting OP of CaFe2As2 as a function of the cell deformation (in the region of the T-cT transition). To avoid the possible complications related to chemical substitutions, we induced this deformation in CaFe2As2 single crystals by means of anisotropic pressure

To investigate the importance of the Lifshitz transition in the superconductivity of IBS and its effects on the symmetry of the OP, we designed a dedicated experimental study of the superconducting OP of CaFe$_2$As$_2$ as a function of the cell deformation induced by non-hydrostatic pressure. To determine the amplitude and structure of the OP, we used an energy-resolved, directional spectroscopic technique such as point-contact Andreev-reflection spectroscopy (PCARS) and interpreted the experimental results with the aid of first-principles Density Functional Theory (DFT) and Random Phase Approximation (RPA) calculations. This combined experimental/theoretical approach shows that the pressure-induced OR-cT transition is accompanied by a topological 2D-3D transition in the holelike FS, the emergence of a horizontal line node in the relevant OP, an increase in $T_c$ and an even greater increase in the OP amplitude that suggests a considerable enhancement of the electron-boson coupling.

\section*{Results}
\subsection*{Point-contact Andreev-reflection spectroscopy}
In a PCARS experiment the differential conductance $dI/dV$ of a nanometric contact between a
normal metal (N) and a superconductor (S) is measured as a function of $V$ ($I$ and $V$ being the probe current through the contact and the voltage at the contact's ends, respectively).
Under suitable conditions (in particular, if the contact diameter is smaller than the electron mean free path \cite{Naidyuk}) the point contact is ballistic and hence spectroscopic, i.e. its differential conductance contains energy-resolved information about the amplitude and the symmetry of the superconducting gap thanks to the quantum phenomenon called Andreev reflection (see Ref. 18 and references therein).
\begin{figure}[ht]
  \vspace{1cm}
  \includegraphics[height=\textwidth,angle=-90]{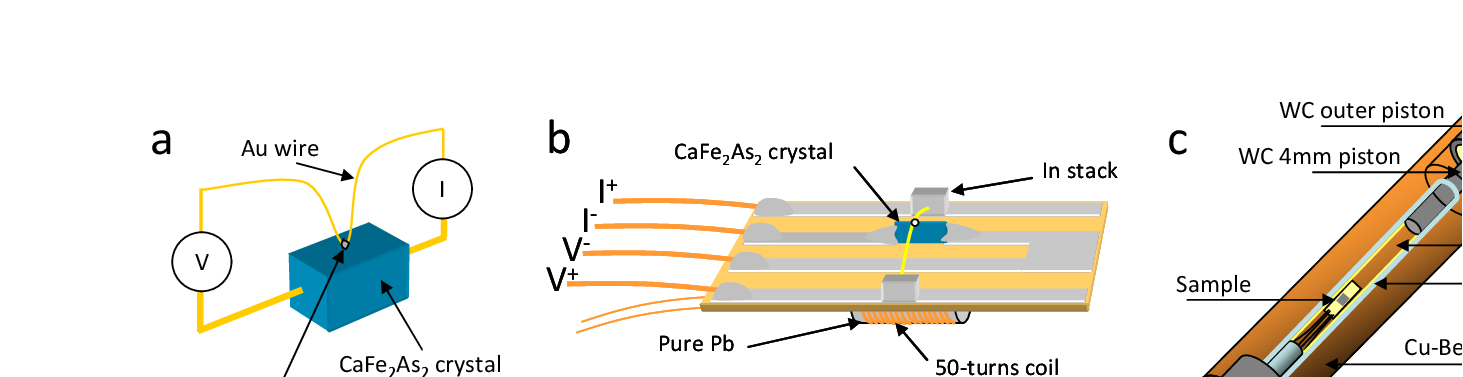}\\
  \vspace{0.5cm}
  \caption{(a) Scheme of the ``soft'' point-contact technique, with the electric connections for PCARS measurements: A probe current $I$ is injected into the crystal through the point contact and the voltage drop $V$  across the contact is measured. (b) Drawing of the fiberglass board used as a sample holder. (c) Drawing of the Cu-Be pressure cell in which the sample is enclosed. }\label{fig:S1}
  \vspace{2mm}
\end{figure}

In the present case, the superconducting samples were CaFe$_2$As$_2$ crystals (of about $2.0\times1.5\times0.1 \,\mathrm{mm}^3$) grown in Sn flux as described in Ref. 19 (see Methods for details). The point contacts were made by putting a small drop of Ag paste ($  \varnothing \simeq  50 \, \mu$m) \cite{Daghero} on the top surface or on the side of the crystals (see Fig.\ref{fig:S1}a) -- which means injecting the current \emph{preferentially} along the $c$ axis or the $ab$ planes, respectively. Within this ``soft'' PCARS technique \cite{Daghero} the contact is a parallel of nanometric junctions that can be destroyed and/or created by short current or voltage pulses \cite{Holm58}. Its resistance can thus be tuned without external mechanical action, so that we could close the crystal in a CuBe clamp cell filled with silicone oil and then pressurized (see Fig.\ref{fig:S1}c). To determine the value of the pressure $P$ (at low $T$) we inductively measured the critical temperature $T_c^{Pb}$ of a small Pb rod (mounted on the sample holder, see Fig.\ref{fig:S1}b) and used the well-known $T_c^{Pb} (P)$ dependence. A uniaxial pressure component $P_{un}$ (essential for superconductivity in CaFe$_2$As$_2$ \cite{Yu,Prokes}) appears when the oil freezes; comparing the pressure dependence of the $T_c$ of our crystals to the analogous curve obtained under uniaxial pressure \cite{Torikachvili09} leads to estimate $P_{un}\simeq 1/4 \,P$.

\begin{figure}[ht]
  \centering
  \includegraphics[height=0.5\textwidth,angle=-90]{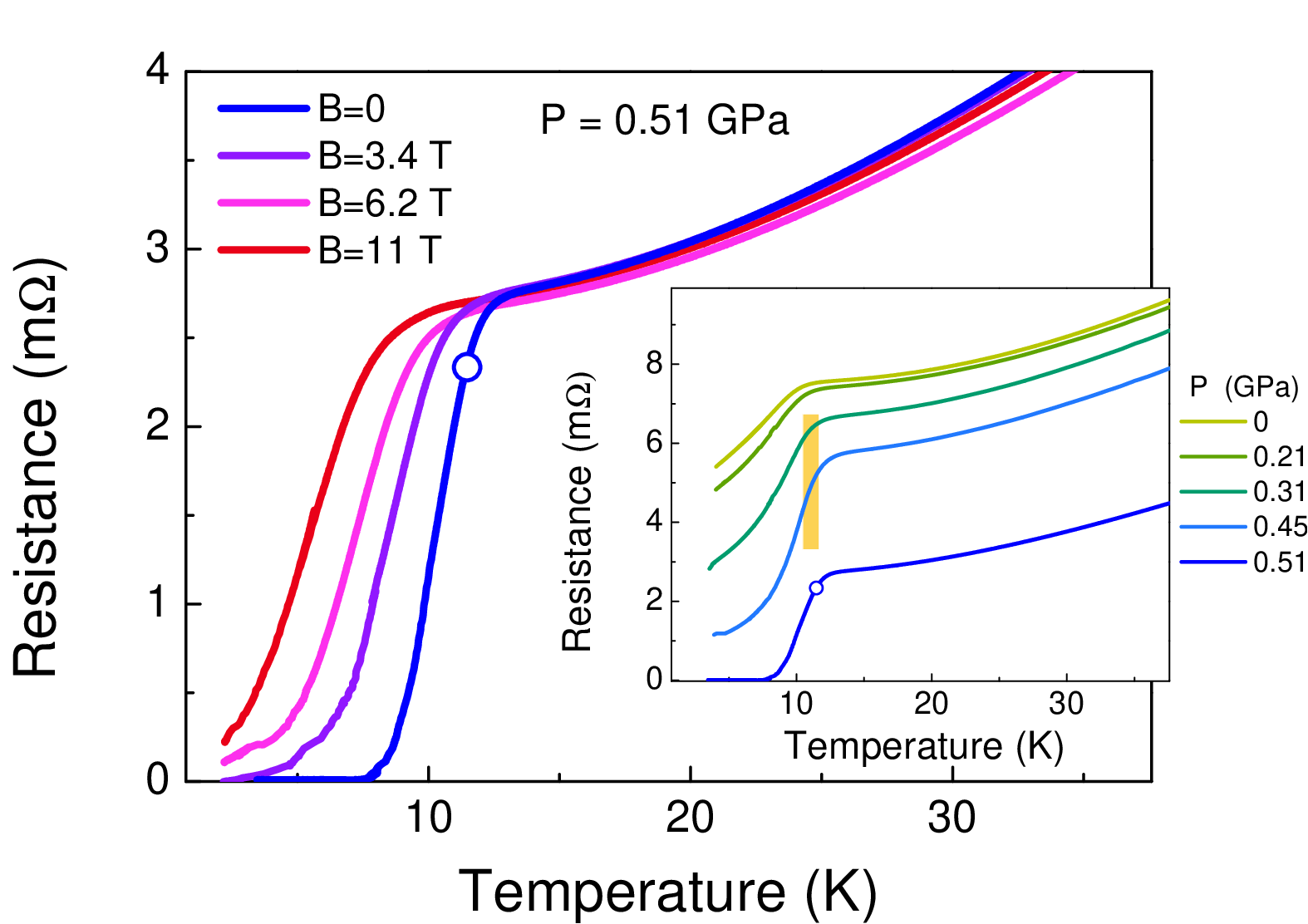}
  %\vspace{-0.2cm}
  \caption{Resistance of a CaFe$_2$As$_2$ crystal under a pressure of 0.51 GPa and in the presence of a magnetic field up to 11 T. Inset: some $R(T)$ curves recorded at different pressures and in zero magnetic field. The curves are vertically offset for clarity. In the main panel, the white dot indicates the values of $T_c^A$ determined by PCARS measurements carried out at 0.51 GPa. In the inset, the range of $T_c^A$ values determined at 0.40 GPa is represented by the width of the colored band behind the curves at 0.31 and 0.45 GPa.}\label{fig:res}
  \vspace{2mm}
\end{figure}

Figure \ref{fig:res} reports the resistance as a function of temperature (in the region of the superconducting transition) of a CaFe$_2$As$_2$ crystal under a pressure of 0.51 GPa and in the presence of a magnetic field up to 11 T. The inset to the same figure reports a subset of $R$ vs. $T$ curves recorded at different pressures from 0 up to 0.51 GPa, zoomed in the region of the superconducting transition. In general, the shape of the $R(T)$ curves up to room temperature is practically identical to that reported by Torikachvili et al. \cite{Torikachvili} apart from the fact that the pressures there are slightly different (indeed, they used a different pressure-transmitting medium, and the uniaxial pressure component was different).

\begin{figure}[ht]
  \centering
  \includegraphics[height=1\textwidth,angle=-90]{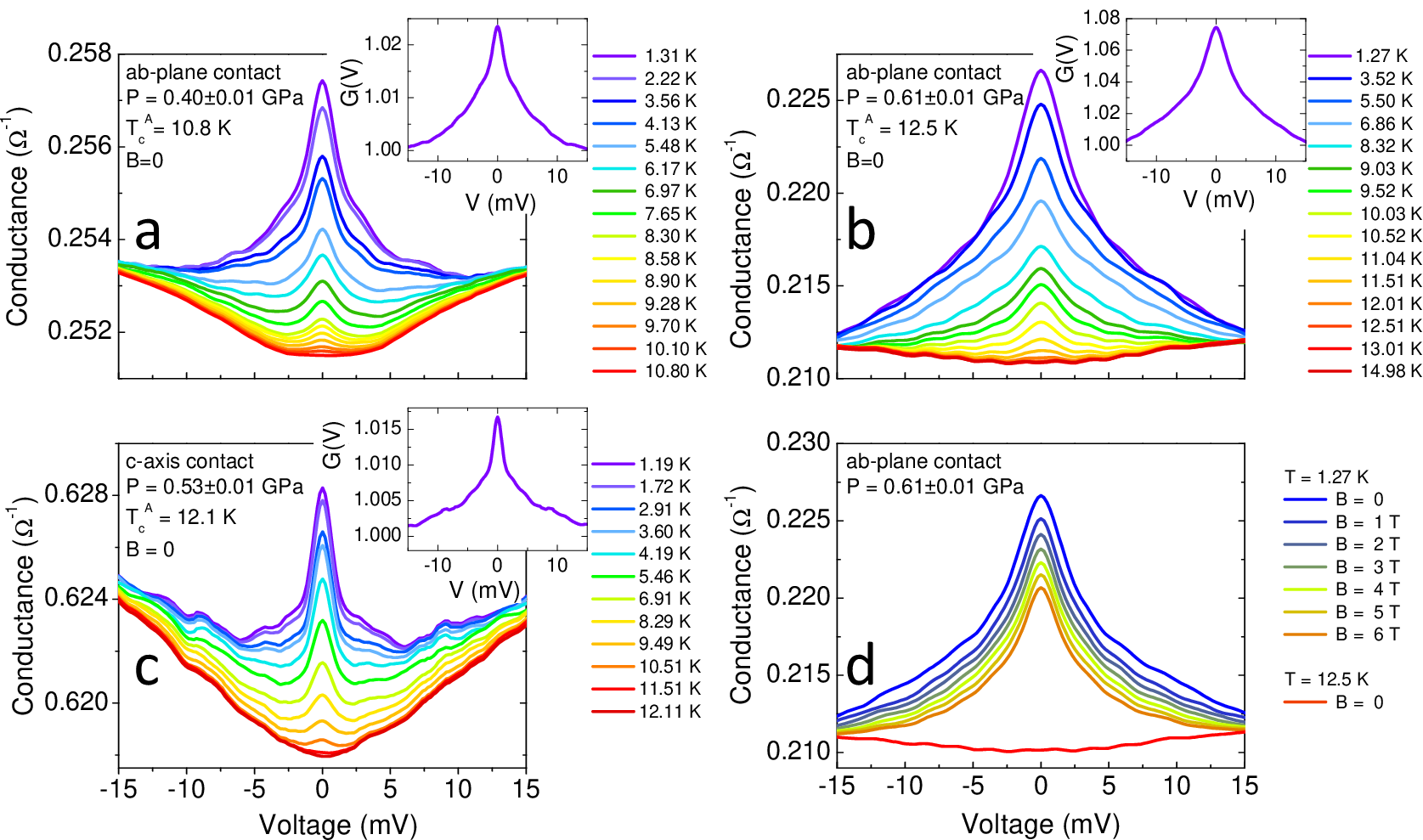}
  \vspace{-0.2cm}
  \caption{(a,b) Raw PCARS spectra of $ab$-plane contacts, measured at various temperatures, in zero magnetic field and at a fixed pressure, i.e. $0.40\pm 0.01$ GPa  and $0.61\pm 0.01$ GPa. (c) Temperature dependence of the raw PCARS spectrum of a $c$-axis contact in zero magnetic field and at $P=0.53\pm 0.01$ GPa. The insets in (a-c) report the normalized low-temperature conductance $G(V)$. (d) Magnetic field dependence (up to $B=6$ T) of the raw differential conductance curve of an $ab$-plane contact at a pressure $P=0.61 \pm 0.01$ GPa and at a temperature $T=1.27$ K. The zero-field curve just above $T_c^A$ (i.e. in the normal state) is also shown for comparison (bottom curve).}\label{fig:1}
  \vspace{-2mm}
\end{figure}

Fig. \ref{fig:1}(a) and \ref{fig:1}(b) show some raw PCARS spectra ($dI/dV$ vs. $V$ curves) measured in ``\emph{ab}-plane contacts'' (i.e. with $I \parallel ab$) at different temperatures and at  $P$=0.40 GPa (a)  and $P$=0.61 GPa (b). The spectra in Fig. \ref{fig:1}(c) were measured in a $c$-axis contact at $P = 0.53$ GPa. The temperature at which the spectra lose any structure and become temperature independent is the ``Andreev critical temperature'' $T_c^A$. Its uncertainty is always of the order of $\pm 0.5$ K. As shown in Fig. \ref{fig:res}, $T_c^A$ always agrees with the critical temperature determined by transport measurements (i.e. it falls on the top of the resistive transition). The PCARS spectrum measured just above $T_c^A$ (i.e. when both banks are in the normal state) will be referred to as $(dI/dV)_{NN}$. The insets to Fig. \ref{fig:1}(a-c) show the low-$T$ spectra \emph{normalized} to $(dI/dV)_{NN}$; in such a way, only the structures associated to superconductivity in the sample are displayed.

Looking at the unnormalized curves, there are various arguments that strongly indicate the contact to be in the ballistic regime and to possess a spectroscopic character \cite{Naidyuk}. First of all, the spectra show no ``dips'' at finite energy \cite{Sheet} that are associated to the quenching of superconductivity in the sample due to overcritical current. The slightly concave shape of the high-voltage tails (even above $T_c^A$) and the good agreement between $T_c^A$ and the bulk $T_c$ determined from resistivity measurements at the same pressure indicate that there is no heating in the contact (i.e. the contact is ballistic). The superposition of the tails on increasing temperature indicates that there is no ``spreading resistance'' in series to the contact \cite{Chen,Daghero,Doring} so that the measured voltage exactly corresponds to the energy of electrons injected in the S bank.

%and that in t the slightly concave shape of the high-voltage tails and their superposition on increasing temperature rule out any factor that would compromise the spectroscopic properties of the contacts , like overcritical current density and Joule heating in the contact region, as well as the existence of a ``spreading resistance'' in series \cite{Chen,Daghero}.
%Also the temperature dependence of the differential conductance can help understanding whether the ZBCP is due to the non-ballistic nature of the contact. In this specific case, the spectra show no sign of Joule heating that would affect their shape on increasing temperature (for example, giving rise to characteristic dips that progressively shift to lower voltage) \cite{RoPP}. The absence of heating effects in the contact is also assured by the good agreement of $T_c^A$ and the bulk $T_c$ determined from resistivity measurements at the same pressure \cite{Torikachvili}.

All the low-$T$ spectra in Fig.\ref{fig:1}(a-c) display a zero-bias conductance peak (ZBCP) and shoulders at 3-5 mV. The same holds for all the spectra we measured, at different pressures and with either $I \parallel ab$ or $I \parallel c$. In general, the ZBCP in PCARS can have various origins: it can be associated to OP anisotropies, in particular involving lines of nodes or zeros \cite{Daghero}, or to spurious effects (contact heating, impurity scattering, etc). It can also arise from Josephson currents if a SIS junction is fortuitously obtained, which is practically impossible in our case due to the ``soft'' PCARS technique we used.

Heating effects can be ruled out since, as previously shown, the contact is ballistic. To see whether the ZBCP could be due to scattering from impurities, we performed model calculations by solving a three-band $s\pm$ version of the Eliashberg equations \cite{eliashberg}, and we found that: a) the impurity level should be extremely high and incompatible with the very high quality of our crystals; b) even in case of such a large amount of impurities, the shape of the ZBCPs would be quite different from the experimental findings.
Further indications about the intrinsic nature of the ZBCP come from the magnetic-field dependence of the PCARS spectra. Fig. \ref{fig:1} d shows the magnetic-field dependence of the low-temperature ($T=1.27$ K) conductance curve of an $ab$-plane contact at a pressure $P=0.61 \pm 0.01$ GPa. The ZBCP is progressively suppressed by the field at approximately the same rate as the shoulders, and this means that both these features are likely to arise from the same mechanism. The ZBCP (and the shoulders) clearly persist up to $B=6$ T. This is evident if the curve at $T=1.27$ K and $B= 6$ T (pink) is compared to the normal-state conductance just above $T_c^A$ (red). The comparison also shows that a field of 6 T is by far smaller than the upper critical field at this temperature and pressure.

%In the present case the ZBCP is intrinsic and not spurious because: i) the contact is ballistic, so that no heating can occur; ii) the ZBCP shows up in 100\% of the contacts; ii) its shape is incompatible with sizable scattering from impurities; iii) it is suppressed by a magnetic field at the same rate as the shoulders, and persists up to $B$=6 T (see Methods for details on the last two points). %; iv) the contacts are spectroscopic.

Therefore, the shape of the spectra in Fig. \ref{fig:1}, as well as their temperature and magnetic-field dependence, strongly suggest the existence of lines of nodes (or zeros) in the OP  \cite{Daghero,Kashiwaya}, but to extract the gap amplitude it is necessary to fit the spectra to a suitable model for Andreev reflection at a NS interface. The simplest possible model that can account for the gap anisotropy is the 2D version \cite{Daghero,Kashiwaya} of the Blonder-Tinkham-Klapwijk  (BTK) model \cite{BTK}. However, this model is unable to fit the spectra if a \emph{single} gap is assumed to exist, independent of its symmetry in the $k$ space. The \emph{two-gap} version of the same model instead allows fitting rather well the $ab$-plane spectra if a large isotropic gap and a small nodal gap are used; unfortunately, in the same conditions it completely fails to fit the $c$-axis spectra. The main reason is that the 2D BTK model is based on the assumption of spherical or cylindrical FS, which is unreasonable for CaFe$_2$As$_2$ \cite{Coldea}. We thus need to use a more refined 3D version of the BTK model \cite{Daghero,RoPP} that can account for any shape of the FS and any $\mathbf{k}$ dependence of the OP. In this model, the \emph{normalized} conductance at $T= 0$ for current injected along the direction $\hat{n}$ is expressed as:

\begin{equation}
%\vspace{-1mm}
\langle G(E) \rangle_{I \parallel n}= \frac{\sum_{i} \langle  \sigma_{i\mathbf{k},n}(E)\tau_{i\mathbf{k},n} \frac{v_{i\mathbf{k},n}}{v_{i\mathbf{k}}} \rangle_{F\!S_i}}{\sum_{i} \langle  \tau_{i\mathbf{k},n} \frac{v_{i\mathbf{k},n}}{v_{i\mathbf{k}}} \rangle_{F\!S_i}} .\label{eq:G}
%\vspace{-1mm}
\end{equation}

Here $\langle \phantom{A}\rangle_{F\!S_i}$ represents the integral over the $i$-th FS sheet; $v_{i\mathbf{k}}$ is the magnitude of the Fermi velocity at $\mathbf{k}$; $v_{i\mathbf{k},n}=\mathbf{v}_{i\mathbf{k}} \cdot \hat{n}$; and $\tau_{i\mathbf{k},n}$ is the normal-state barrier transparency which depends on the Fermi velocities and on the potential barrier at the NS interface (represented by a dimensionless parameter $Z_i$). Finally $\sigma_{i\mathbf{k},n}(E)$ is the relative barrier transparency in the superconducting state, that contains the information about the $\mathbf{k}$-dependence of the OP on the $i$-th FS sheet \cite{Daghero,RoPP}. The normalized conductance at any $T\neq 0$ can be calculated by convolution of $\langle G(E) \rangle_{I \parallel n}$ with the Fermi function.
%{\color{blue}{Note that the knowledge of the shape of the Fermi surface is necessary in order to use eq. \ref{eq:G} to fit the experimental PCARS spectra. In principle, the $k$ dependence of the gap should be know as well; if it is not, one can make a guess and see whether it allows fitting the experimental data.}}

\subsection*{Calculation of the Fermi surface}
\begin{figure}[ht]
  \centering
  \vspace{10mm}
  \includegraphics[height=0.5\columnwidth,angle=-90]{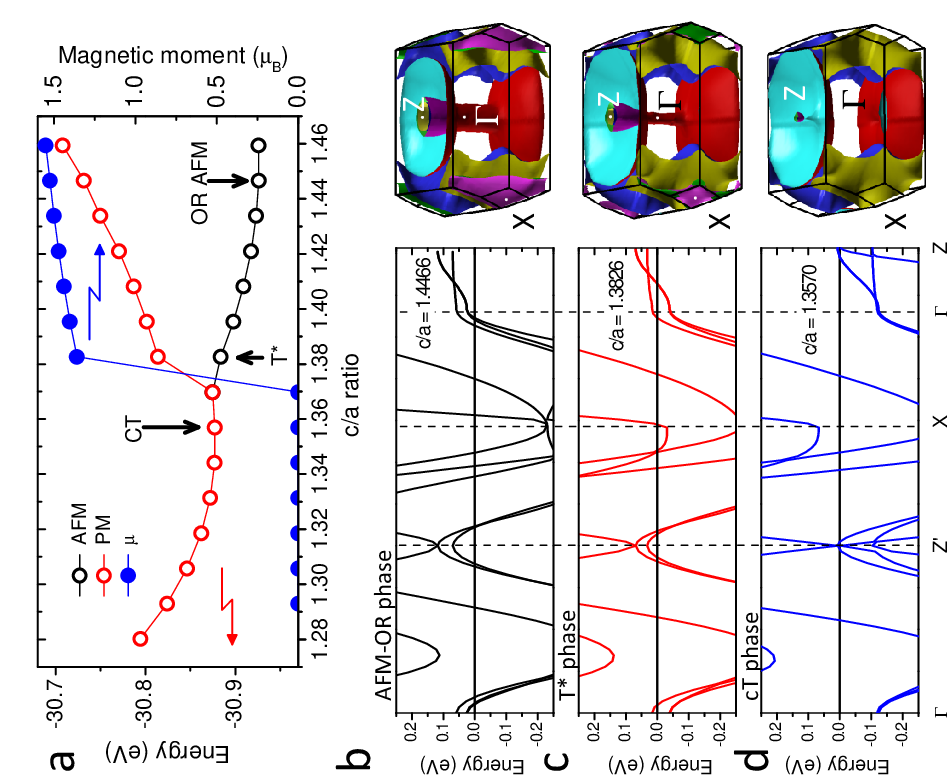}
  %\end{center}
  %\vspace{-0.2cm}
  \caption{(a) Fe magnetic moment (solid circles) and total energy in the AFM (black circles) and PM (red circles) phases, as a function of $c/a$. The $c/a$ values corresponding to the OR, cT and T* structures are indicated by arrows. %Inset: the parameter $h_{As}$ and the Fe-As-Fe bond angle versus $c/a$.
  (b-d) Electronic band structure and Fermi surface in the OR AFM (b), T* (c) and cT structure (d). }\label{fig:2}
  %\vspace{-2mm}
\end{figure}

Since the 3D BTK model requires knowing the shape of the FS, we calculated the band structure of CaFe$_2$As$_2$ by DFT using both pseudopotential \cite{pwscf} and all-electron full-potential \cite{wien2k,elk} codes,  within the generalized gradient approximation (GGA) for the exchange-correlation functional, using the PBE approach \cite{Perdew}  (see Methods for details). Uniaxial pressure was simulated by reducing the lattice parameter $c$ and leaving the in-plane one $a$ fixed to its equilibrium value at $P=0$ ($a= 3.925079$ {\AA}). This is justified by the fact that, up to 6 GPa, $a$ decreases by about 1\%, while  $c$ by about 10\% \cite{Colonna}. For any $c/a$ value, the equilibrium structure corresponds to the value of $h_{As}$ (the height of the As atoms above the Fe layer) that minimizes the total energy -- determined considering the AFM (PM) phase for the orthorhombic (tetragonal) structure \cite{Sanna}. Fig. \ref{fig:2}(a) reports the total energy and the magnetic moment $\mu$ in the AFM and PM phases, versus the $c/a$ ratio. The structure at $P$= 0 ($c/a = 1.4466$, $\mu=  1.5$ $\mu_{B}$) corresponds to the minimum energy in the AFM OR phase. On decreasing $c/a$ the energy of this phase increases and, for $c/a < 1.3826$, the system undergoes a sharp first-order transition towards a non-magnetic phase. We associate the narrow region around this $c/a$ value to the non-collapsed tetragonal phase (T*) where superconductivity is thought to exist \cite{Prokes}. Note that at the onset of the phase transition  $h_{As}$ and the As-Fe-As bond angle jump to values close to the ``optimal'' ones observed at the maximum $T_c$ for various Fe-based compounds \cite{Johnston}. % (inset to Fig. \ref{fig:2}(a)).
On further increasing pressure, the total energy exhibits a second minimum at $c/a= 1.3570$, related to the cT phase -- this $c/a$ value being in excellent agreement with that calculated under hydrostatic conditions \cite{Colonna}.
\\

Fig.~\ref{fig:2} also shows the electronic structure and the FS of CaFe$_2$As$_2$ in the equilibrium structure at $P$= 0 (b), in the T* phase at the edge of the phase transition (c), and in the equilibrium cT structure (d). The FS at $P$=0 consists of three holelike sheets (two almost cylindrical, one largely warped at $k_z=\pi$) centered at the $\Gamma$  point of the Brillouin zone (BZ) and two 2D electronlike sheets centered at the $X$ point. In the T* phase ($c/a= 1.3826$)  the two inner hole bands are completely filled at $\Gamma$ and give rise to two 3D Fermi pockets centered at the $Z$ points. The outer holelike FS further expands at $k_z=\pi$ and shrinks at $k_z=0$. Finally, in the cT phase, where the Fe magnetic moment vanishes, the inner hole bands are filled both at $\Gamma$ and $Z$, and the outer hole band exhibits full 3D character, crossing the Fermi level along the $\Gamma-Z$ line. These latter findings are in agreement with previous calculations \cite{Tompsett}.
The T* phase is thus on the verge of a topological 2D-3D transition of the holelike FS but still retains a small FS at $\Gamma$ and a nonzero magnetic moment, and is characterized by almost ``optimal'' values of $h_{As}$ and of the As-Fe-As angle. % (inset to Fig. \ref{fig:2}(a)).
We can thus assume this phase to be (close to) the one that displays optimal superconducting properties.

\subsection*{Calculation of the order parameter structure}
In addition to the shape of the FS, the 3D BTK model also requires the knowledge of the $k$ dependence of the OP. Even in the absence of any specific information about this point, some simulations with the 3D BTK model allow drawing some qualitative conclusions. Indeed, to reproduce the existence of a zero-bias maximum in \emph{both} $ab$-plane and $c$-axis spectra, i) the shoulders must be associated to a ``large'' gap, possibly isotropic, and the zero-bias maximum to a smaller gap, featuring lines of nodes or zeros; ii) this small gap must reside on a FS with a pronounced 3D character (i.e. not on a hyperboloid-like FS sheet, as shown elsewhere \cite{RoPP}; iii) the lines of nodes (or zeros) can be either vertical or horizontal (for further details see Supplementary Information). These conclusions actually agree with what we found in 6\% Co-doped CaFe$_2$As$_2$ \cite{GonnelliCaFeCoAs,low_temp_phys} (that, by the way, has a Fermi surface similar to that of the T* structure shown in Fig. \ref{fig:2}c); theoretical predictions for the the OP structure in the 122 family of IBS \cite{Suzuki,Graser,Hirschfeld} indicate that the lines of nodes on the holelike FS may be horizontal rather than vertical.

\begin{figure}[ht]
  \centering
  \includegraphics[width=0.5\columnwidth]{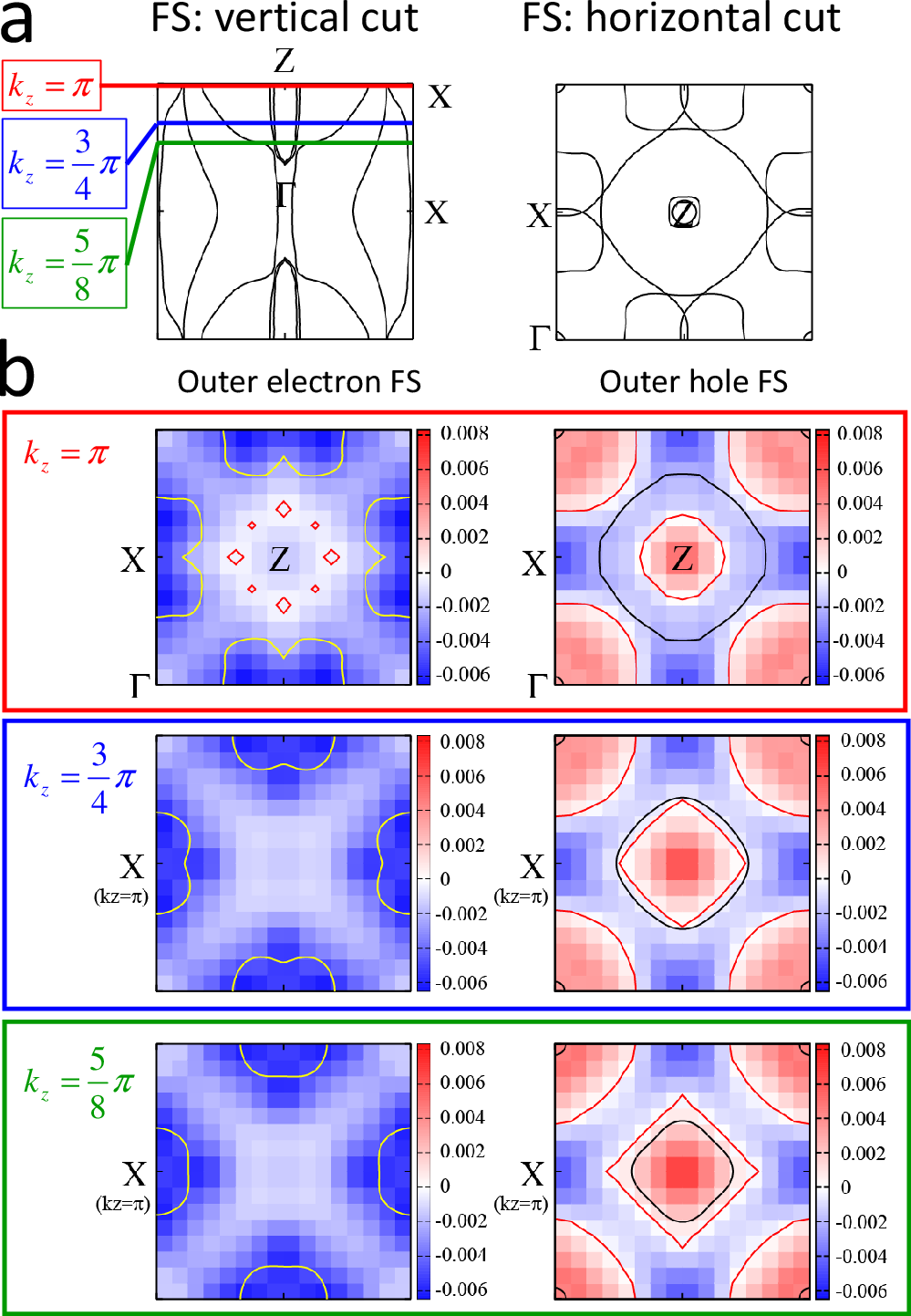}
  %\end{center}
  \vspace{1cm}
  \caption{(a) Vertical cut and horizontal cut (at $k_z = \pi$) of the FS of the T* structure. (b) Calculated superconducting OP at (from top to bottom) $k_z=\pi$, $3\pi/4$ and  $5 \pi/8$ for the outer electronlike FS sheets (left) and the outer holelike FS sheet (right). The profile of the electronlike (holelike) FS is indicated by yellow (black) lines; red lines indicate instead the node lines. The gap amplitude is given by the color scale.}\label{fig:3}
  %\vspace{-2mm}
\end{figure}

To get some more specific hints about the $k$ dependence of the OP in the T* structure of CaFe$_2$As$_2$, we followed the method adopted for BaFe$_2$(As,P)$_2$ \cite{Suzuki}: Starting from the band structure in the T* phase we constructed a ten-orbital model (five 3d orbitals $\times$ two Fe atoms per unit cell) exploiting the maximally localized Wannier functions \cite{Marzari,wannier}. We then applied a Random Phase Approximation (RPA) approach to obtain the spin susceptibility, and we solved the linearized Eliashberg equations using a pairing interaction proportional to this spin susceptibility (see Methods for details). The superconducting gaps for the quasi-2D outer electron sheet and the quasi-3D outer hole sheet are depicted in Fig. \ref{fig:3} at different $k_z$ values. The OP on the electron FS, $\Delta_e$, is relatively isotropic (except at $k_z=\pi$), while the OP on the hole FS, $\Delta_h$, is smaller and strongly anisotropic in the $ab$ plane. Moreover, the latter decreases on decreasing $k_z$ until it changes sign at $k_z \simeq 3 \pi /4$, forming a horizontal node line.
These calculations thus confirm the qualitative conclusions about the OP structure drawn on the sole basis of the shape of the PCARS spectra, and give them a robust foundation; however, they do not provide reliable values of the gap amplitudes $\Delta_e$ and $\Delta_h$. These quantities can be obtained by fitting the PCARS spectra with the 3D BTK model (eq.\ref{eq:G}).

\begin{figure}[ht]
  \centering
  \vspace{5mm}
  \includegraphics[height=0.5\columnwidth,angle=-90]{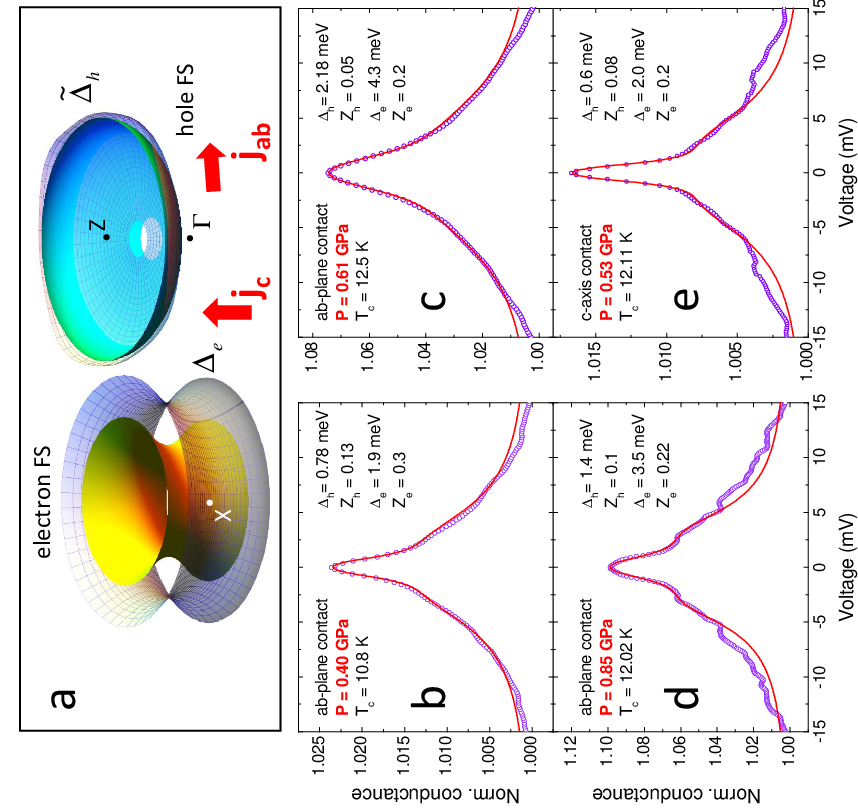}
  %\end{center}
  %\vspace{-0.2cm}
  \caption{(a) Model FS used in the 3D BTK model (matt surfaces) in the upper half of the BZ. Gridded surfaces represent the superconducting OP. %The high-symmetry points are indicated.
  (b-d) Low-temperature PCARS spectra (symbols) of $ab$-plane contacts at $P$= 0.4 GPa (b), 0.61 GPa (c) and 0.85 GPa (d) with the relevant 3D BTK fit (lines). (e) Low-temperature spectrum of a $c$-axis contact at $P = 0.53$ GPa, with the relevant fit. The fitting parameters are indicated in the legends.}\label{fig:4}
  %\vspace{-2mm}
\end{figure}

\subsection*{Fit of the PCARS spectra}
For ease of calculation, we schematize the outer electronlike FS in the upper half of the BZ by a one-sheeted hyperboloid of revolution and the outer holelike FS by an oblate spheroid centered at $Z$ (see Fig. \ref{fig:4}(a)). This simplification preserves all the important characteristics of the FS from the point of view of eq.\ref{eq:G}. For example, the cylindrical neck along $\Gamma-Z$ (see fig. \ref{fig:2}(c)) gives a small (for $I \parallel ab$) or null (for $I \parallel c$) contribution to the conductance and can be neglected, leaving a small hole in the spheroid. According to the results of RPA calculations, the gap on the electronlike FS ($\Delta_e$) is isotropic, but that on the holelike FS has a horizontal node line at $k_z = 3\pi/4$ and a fourfold in-plane anisotropy (see Fig.\ref{fig:4}(a)). $\Delta_h$ is defined so that at $k_z=\pi$ the holelike gap varies between 0.5$\Delta_h$ and 1.5$\Delta_h$.

The adjustable parameters of the 3D BTK model are thus the gap amplitudes $\Delta_e$ and $\Delta_h$, the barrier parameters $Z_e$ and $Z_h$, and the broadening parameters $\Gamma_e$ and $\Gamma_h$ that account for the small amplitude of the Andreev signal \cite{RoPP}. The weight of each band is fixed by the FS geometry and the direction of current injection.
The results of the best fits --always obtained by minimizing the sum of squared residuals-- are shown in Fig. \ref{fig:4}(b-d) for $ab$-plane contacts (at $P$= 0.40, 0.61 and 0.85 GPa) and in \ref{fig:4}(e) for a $c$-axis contact (at 0.53 GPa). The fit is very good for both the injected-current directions, and similar results have been obtained for all the contacts at different pressures. Note that the OP structure indicated by RPA calculations allows the best fit of the spectra, if compared to other configurations (e.g. vertical node lines on the holelike FS). Further details on this point as well as on the error bars and the robustness of the fits can be found in Supplementary Information.

\begin{figure}[ht]
  \centering
  \includegraphics[height=0.6\columnwidth,angle=-90]{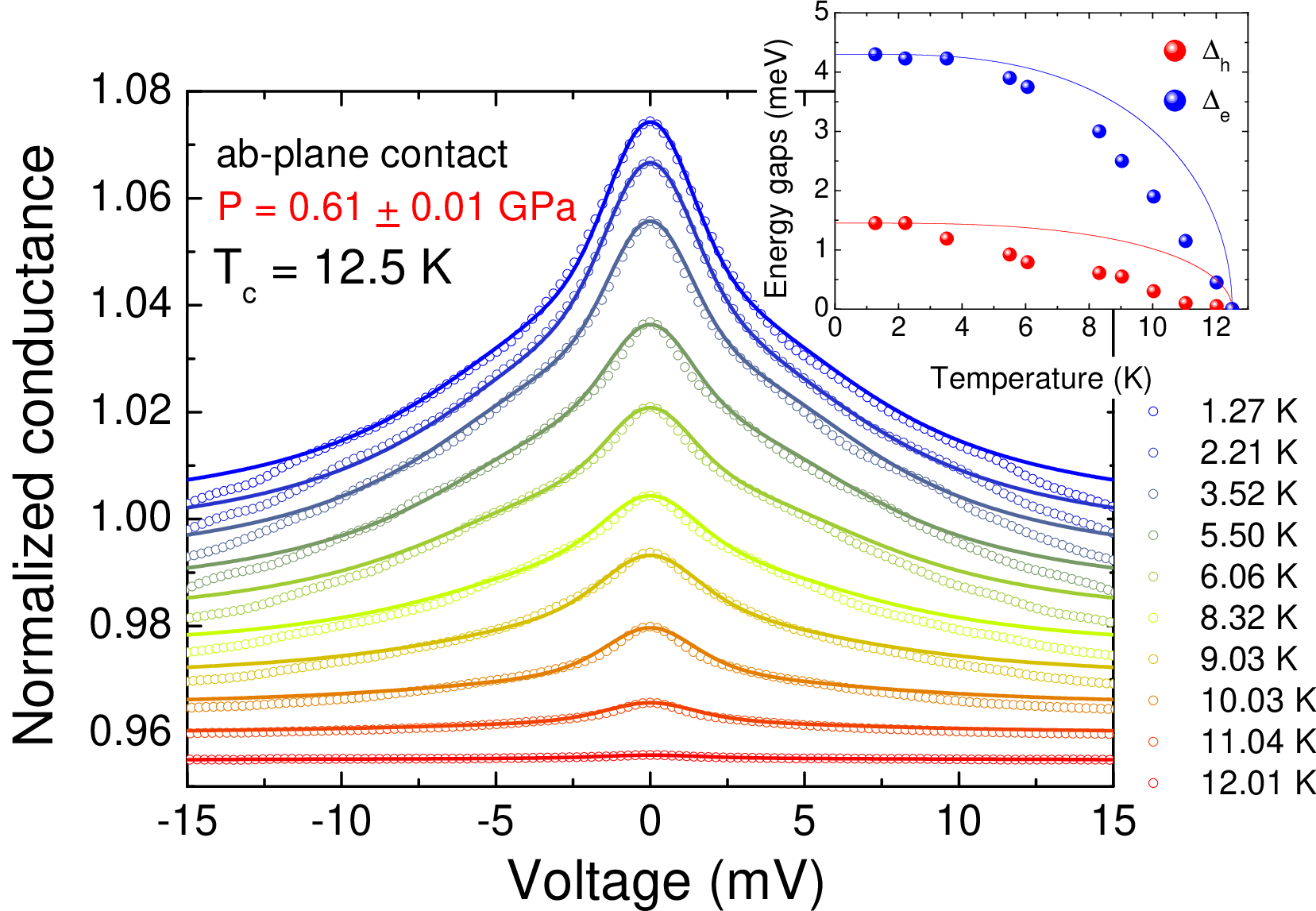}
  %\end{center}
  %\vspace{-0.2cm}
  \caption{The normalized PCARS spectra of an $ab$-plane contact at different temperatures between 1.27 K and 12.01 K (symbols) and the relevant fit (lines) obtained within the 3D BTK model. All the curves apart from the lowest-temperature one have been vertically shifted for clarity. The inset shows the temperature dependence of the amplitudes of the gaps extracted from the fit: the nodeless gap on the electron-like FS sheet, $\Delta_e$ (blue symbols) and the nodal gap $\Delta_h$ on the hole-like FS sheet (red symbols). BCS-like $\Delta(T)$ curves are shown for the sake of comparison (lines).}\label{fig:S3}
  \vspace{2mm}
\end{figure}

Once the spectra measured at different temperatures are normalized, it is possible to fit them with the 3D BTK model to determine how the amplitude of the OPs depends on temperature.
Fig. \ref{fig:S3} reports, as an example, a set of conductance curves of the same $ab$-plane contact at $P = 0.61 \pm 0.01$ GPa  measured at different temperatures between 1.27 and 12.01 K (symbols). The 3D BTK fits (solid lines) of the curves are shown as well. The fit allowed us to extract the temperature dependence of the gap amplitudes $\Delta_h$ and $\Delta_e$ that is reported in the inset. Similar results are obtained from the temperature dependencies of the normalized PCARS spectra at the other pressures.

\subsection*{Pressure dependence of the gap amplitudes}
\begin{figure}[ht]
  \centering
  \vspace{-2cm}
  \includegraphics[width=0.5\columnwidth]{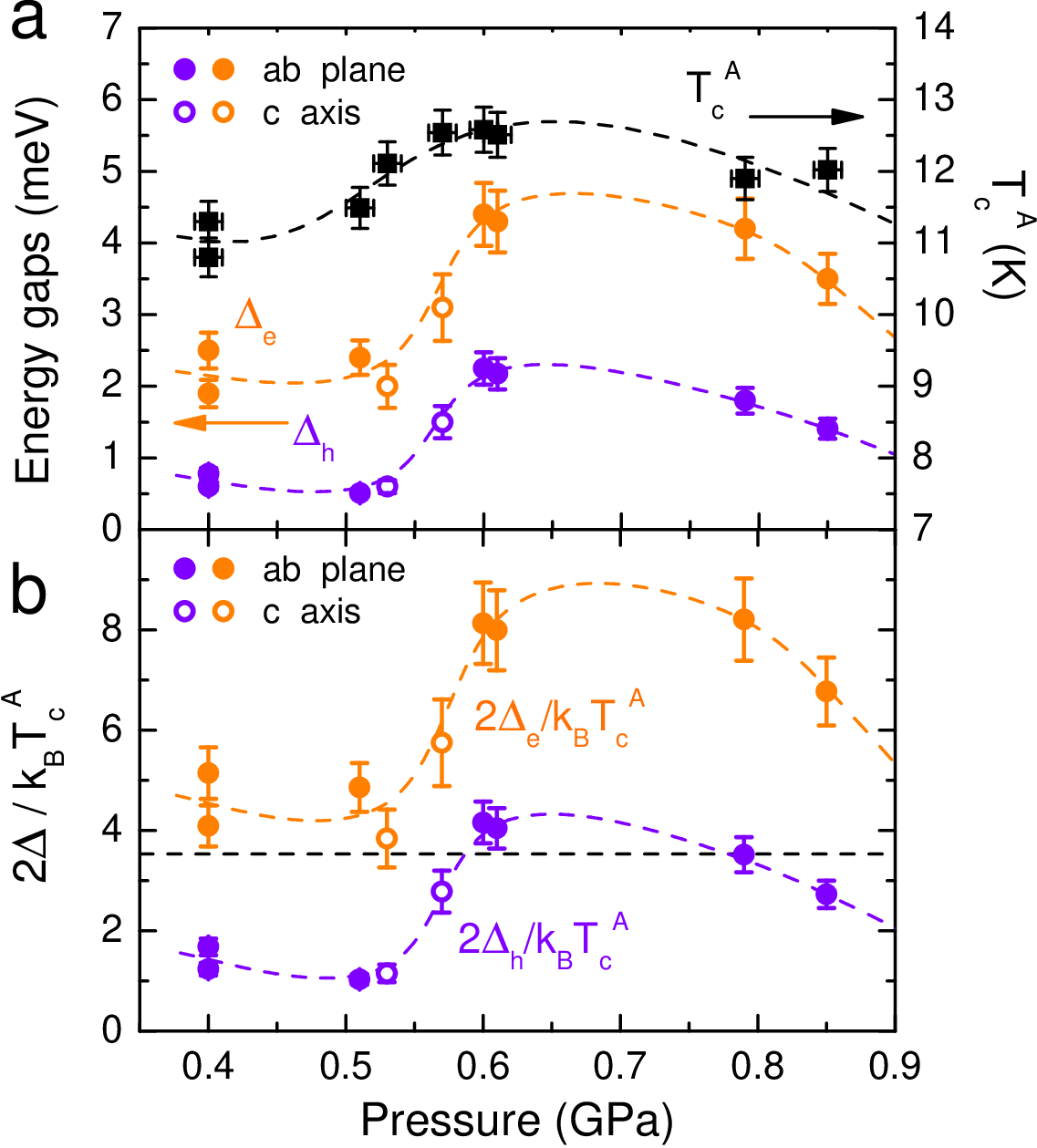}
  % \end{center}
  \vspace{3cm}
  \caption{(a) $T_c^A$ (black squares, right vertical scale) and gap amplitudes $\Delta_e$ and $\Delta_h$ (orange and purple circles, left vertical scale), as a function of pressure. (b) Pressure dependence of $2\Delta_e/k_BT_c$  (orange circles) and  $2\Delta_h/k_BT_c$ (purple circles). In (a) and (b) data from $ab$-plane and $c$-axis contacts are shown as solid and open symbols, respectively.} \label{fig:5}
\end{figure}

Fig. \ref{fig:5}(a) reports $T_c^A$, $\Delta_e$ and $\Delta_h$ vs. pressure. The error bars on the gaps represent the range of values that give an acceptable fit (within a fixed confidence level) when all the other parameters are changed as well (see Supplementary Information for the procedure). $T_c^A$ follows very well the bulk $T_c$ reported elsewhere \cite{Torikachvili}, showing a broad maximum between 0.6 and 0.75 GPa. The gaps remain almost constant between 0.4 and 0.55 GPa, but between 0.55 and 0.6 GPa they increase so much that even the gap ratios $2\Delta_h/k_BT_c$ and $2\Delta_e/k_BT_c$ grow (see Fig. \ref{fig:5}(b)) becoming both greater than the BCS value 3.52; $2\Delta_e/k_BT_c$ even reaches values of about 8. Note that similar gap ratios were obtained in 6\% Co-doped CaFe$_2$As$_2$, which is close to an analogous Lifshitz transition induced by chemical doping \cite{low_temp_phys}. In KFe$_2$As$_2$ (that shows no electrolike FS) gap ratios equal to $7.58 \pm 0.34$, $2.80 \pm 0.34$ and $1.02 \pm 0.41$ were measured by ARPES on the FS sheets around $\Gamma$ \cite{okazaki}. Even larger values were found in (Ba$_{0.1}$K$_{0.9}$)Fe$_2$As$_2$ \cite{Xu}.

\section*{Discussion}
The enhancement of the critical temperature and of the gap amplitudes in a narrow pressure range, and even more the steep increase in the gap ratios, suggest an enhancement in the electron-boson coupling in concomitance with the 2D-3D topological transition, which can still be explained within a spin-fluctuation mediated pairing (even though alternative strong-coupling explanations \cite{Xu} cannot be excluded): When the holelike FS shrinks, the \textit{set} of ``nesting wavevectors'' that connect it to \textit{portions} of the electronlike FS widens, and the integral of the susceptibility increases -- until the holelike FS completely vanishes at $\Gamma$ and superconductivity disappears. The occurrence of a maximum in $T_c$ and in the gap amplitudes in correspondence of a 2D-3D Lifshitz transition has been recently predicted -- within a BCS approach, with the inclusion of the density equation -- in multiband superconductors when the chemical potential is tuned near a band edge \cite{Innocenti}. This picture holds for 3D superlattices of quantum wells, of which IBS, among other layered superconductors, can be considered a practical realization, and is based on the concept of shape resonance. In similar conditions, a peak in the isotope coefficient has also been predicted \cite{Perali}.

To summarize the results obtained by the unique combination of directional PCARS on CaFe$_2$As$_2$ under quasi-hydrostatic pressure and state-of-the-art ab-initio calculations proposed in the present paper, it can be said that: i) the electron-boson coupling largely increases in the narrow pressure range where a topological 2D-3D transition of the holelike FS occurs; ii) this transition drives the emergence of a horizontal node line in the OP on the holelike FS sheet, which leaves its fingerprints in the zero-bias Andreev-reflection conductance.
It is intriguing to notice that the same phase transition with disappearance of the holelike FS at $\Gamma$  that here produces a dramatic increase of the coupling in both bands seems to be responsible for the high-$T_c$ phase at 47 K observed in rare-earth doped CaFe$_2$As$_2$ \cite{Saha}, and for the second, higher-$T_c$ superconducting dome in LaFeAsO$_{1-x}$H$_x$ \cite{Iimura}. Recently, a study of KFe$_2$As$_2$ in an extended pressure range (up to 30 GPa) has shown \emph{two} superconducting phases in the phase diagram \cite{Nakajima}. A first superconducting dome (with a maximum $T_c \simeq 7$ K around 2.5 GPa) occurs in the T structure, and is extended up to about 7 GPa. Note that, as previously reported, no electronlike FS sheets exist in this structure. A second superconducting phase with a greatly enhanced critical temperature (about 11 K) appears abruptly when the system undergoes the transition to the cT phase at $P= 13$ GPa; $T_c$ then decreases on further increasing the pressure. The sudden jump in $T_c$ around 13 GPa is ascribed to the fact that the holelike FS sheet at $\Gamma$ shrink, the holelike FS at $X$ disappears and is replaced by an electronlike pocket, thus restoring the usual conditions of imperfect nesting associated to the spin-fluctuation pairing mechanism. In this case, as in ours, the maximum $T_c$ occurs exactly at the T-cT transition.

%This approach can provide a new comprehensive tool for the fundamental investigation of the effect on superconductivity of subtle changes in the Fermi-surface topology.

%\section*{Discussion}
%
%The Discussion should be succinct and must not contain subheadings.

\section*{Methods}
\subsection*{Crystal growth}
The CaFe$_2$As$_2$ single crystals were grown in a Sn flux, starting from atomic ratios Ca:Fe:As:Sn=1.1:2:2.1:40. The components, contained in alumina crucibles, were placed in sealed quartz ampoules previously filled with Ar gas under a pressure of 0.3 bar. The ampoules were then slowly heated ($\sim  4$ h) to 600 $^{\circ}$C, held at this temperature for 1 h, then further heated ( $\sim  5$ h) to 1050 $^{\circ}$C, and kept at this temperature for 5 h so that all the components dissolved in the Sn flux. Crystals were grown by slow cooling the melt at a rate of 2 K/h to 600 $^{\circ}$C, followed by furnace cooling to room temperature. The flux was finally removed by etching in diluted hydrochloric acid.
The resulting single crystals were platelike, with the c-axis perpendicular to the plate. Phase purity was checked by powder x-ray diffraction (XRD). The details of the characterization can be found in Ref. \cite{Matusiak}.

\subsection*{Experimental setup for point-contact measurements under pressure}
The most usual way for making the point contact in PCARS is the so-called needle-anvil technique, in which a sharp normal-metal tip is pressed against the surface of the superconductor \cite{Naidyuk}. The resistance of the contact is generally controlled by moving the tip forward/ backward through a rigid shaft connected to a suitable device (i.e. a stepper motor). This setup is obviously not applicable if the sample has to be closed in a high-pressure cell, as in our case.
We thus adopted an alternative setup (the ``soft'' point-contact technique we introduced some years ago \cite{Daghero}) wherein the contact tip is substituted by a tiny silver paste contact whose resistance can be tuned via short current or voltage pulses. This method can be easily adapted to work in a pressure cell since it allows the whole setup to be mounted on a small support, and does not require any mechanical action from external moving parts. Moreover, it ensures a greater mechanical, thermal and electrical stability of the contacts.

In this specific case, the small drop of Ag paste ($\varnothing \simeq  50 \, \mu$m) was put on the freshly cleaved surface of the crystal. A varying probe current $I$ was then injected from the Ag spot into the superconductor and the voltage $V$ across the contact was measured, thus determining the $I-V$ characteristics of the contact (see Fig. \ref{fig:S1}(a)) and then the differential conductance $dI/dV$ vs. $V$. Note that if the contact is put on the topmost surface, the current is mainly injected along the $c$ axis; if it is placed on the side surface, the current is preferentially injected along the $ab$ planes.

The sample holder was made from a small ($\simeq  7\times 3 \, \mathrm{mm}^2$) fiberglass board, with a small 50-turn coil wrapped around a piece of high-purity Pb (99.9999\%) placed on the reverse side (see Fig. \ref{fig:S1}(b)). The sample holder was mounted in a Cu-Be pressure cell and immersed in silicon oil used as the pressure-transmitting medium. This ensures hydrostatic pressure conditions above the freezing point of the oil, but when the oil freezes a uniaxial pressure component sets in. The pressure was applied at room temperature, then the cell was cooled in a $^4$He variable temperature cryostat. To determine the pressure at low temperature, we measured the temperature of the superconducting transition of Pb by an ac mutual inductance technique; then we used the well-known $T_c(P)$  dependence for Pb to extract the pressure value. This technique allows determining the pressure at low temperature, but does not allow separating the uniaxial component from the hydrostatic one; therefore, a scaling factor might be necessary when comparing our results with those obtained by using different pressure-transmitting media.

\subsection*{DFT and RPA ab-initio calculations}
The ab-initio DFT calculations were performed by using both the pseudopotential method (Vienna ab-initio Simulation Package, Quantum Espresso \cite{pwscf}) and all-electron full-potential codes (Wien2K \cite{wien2k}, Elk \cite{elk}) in order to compare the results and confirm the findings.
For the structural optimization (always done in the antiferromagnetic phase) the charge density was integrated with $ 8 \times 8 \times 4$ $k$-points in the Brillouin zone. The band structures were always calculated in the non-magnetic body-centered tetragonal phase (irrespective of their ground state).
%Figure \ref{fig:bands} reports the band structure for the OR, T* and cT phases, on a larger energy scale than figure 2b.
%
%\begin{figure}[h]
%  % Requires \usepackage{graphicx}
%  \includegraphics[width=0.5\columnwidth]{band_ext.eps}\\
%  %\vspace{-0.2cm}
%  \caption{\small{Band structure of the OR (top), T* (middle) and cT phase (bottom) on an expanded energy scale, i.e. from $-2.0$ eV to $+ 2.0$ eV around the Fermi level.}}\label{fig:bands}
%  \vspace{2mm}
%\end{figure}
%
For the construction of the ten-orbital model that takes into account all five 3d orbitals we exploited the maximally localized Wanner orbitals \cite{Marzari} by using the code developed by A. A. Mostofi et al. for the generation of the Wannier functions \cite{wannier} starting from first-principles band structure. As for the electron-electron interactions, we considered the intra-orbital $U=1.56$, the inter-orbital $U'=1.17$, Hund's coupling and the pair hopping interaction $J=J'=0.195$. In the RPA calculations used to obtain the spin susceptibility on $16 \times 16 \times 16$ $k$-point meshes, 512 Matsubara frequencies were used and the temperature was fixed at $T=0.04\, eV$. The calculation of the OP symmetry has been carried out for one specific value of the $c/a$ ratio (1.3826, corresponding to the T* structure) and thus one pressure (i.e. $P \sim  0.5$ GPa) but qualitatively holds also for the other pressures used in the PCARS experiments, as long as the XZ and YZ hole bands, which provide the largest contribution to the DOS, are filled at $\Gamma$  while the outer holelike FS survives.
The complex structure of the order parameters, obtained by solving the linearized Eliashberg equation using the pairing interaction proportional to the RPA spin susceptibility, was used to fit the PCARS spectra with the 3D BTK model. Actually, to simplify the calculations, both the FS and the OP were modeled by analytical functions in spherical coordinates. In particular, the $\mathbf{k}$-dependence of the OP on the holelike FS was approximated by the following function:

%\begin{dmath}
%  Q(\lambda,\hat{\lambda}) = -\frac{1}{2} P{(O \mid \lambda )} \sum_s \sum_m \sum_t \gamma_m^{(s)} (t) \left( n \log(2 \pi ) + \log \left| C_m^{(s)} \right| + \left( \mathbf{o}_t - \hat{\mu}_m^{(s)} \right) ^T C_m^{(s)-1} \left(\mathbf{o}_t - \hat{\mu}_m^{(s)}\right) \right)
%\end{dmath}
%
\begin{equation*}
\widetilde{\Delta}_h(\theta, \phi) = 2 \Delta_h \left[ \sin\left(\phi + \frac{\pi}{3}\right)
\Theta\left(-\phi + \frac{2 \pi}{3}\right)\left[\cos^4(2\theta)+\frac{1}{2} \right]
 + \sin\left(\phi + \frac{\pi}{3}\right)\Theta\left(\phi - \frac{2 \pi}{3}\right) \right]
\end{equation*}
where $\theta$ and $\phi$  are the azimuth and the inclination angles in the $k$ space, respectively, and $\Theta$  is the Heaviside step function. In the $k_z=\pi$ plane ($\phi=\pi/2$), the gap amplitude varies (depending on $\theta$) between $0.5 \Delta_h$ and $1.5 \Delta_h$  (note that $\phi=\pi$ corresponds to the $Z-\Gamma$ direction). The $\widetilde{\Delta}_h(\theta, \phi)$  function, together with the constant order parameter on the electronlike FS  $\Delta_e$, is plotted in Fig. 4(a) of the main text (gridded surfaces) and is used in the 3D BTK fit of the experimental data.

%\subsection*{Check of the intrinsic origin of the zero-bias conductance peak}
%When the conductance curve of a point contact displays a zero-bias conductance peak (ZBCP), it is necessary to clarify whether this peak is intrinsic (i.e. due to Andreev reflection) or instead arises from spurious effect.
%
%For example, in order to investigate whether the ZBCP could be due to scattering from impurities, we performed model calculations by solving a three-band $s\pm$ version of the Eliashberg equations, and we found that: a) the impurity level should be extremely high and incompatible with the very high quality of our crystals; b) even in case of such a large amount of impurities, the shape of the ZBCPs would be quite different from the experimental findings.

%

%Also the temperature dependence of the differential conductance can help understanding whether the ZBCP is due to the non-ballistic nature of the contact. In this specific case, the spectra show no sign of Joule heating that would affect their shape on increasing temperature (for example, giving rise to characteristic dips that progressively shift to lower voltage) \cite{RoPP}. The absence of heating effects in the contact is also assured by the good agreement of $T_c^A$ and the bulk $T_c$ determined from resistivity measurements at the same pressure \cite{Torikachvili}.

%\end{spacing}
%\section*{References}

\section*{Acknowledgments}
This work was supported by the Collaborative EU-Japan Project ``IRON SEA'' (NMP3-SL-2011-283141).
R.S.G. wishes to thank the Max-Planck Institute for Solid State Research in Stuttgart where the measurements were carried out.

\section*{Author contributions statement}
R.S.G. designed the PCARS measurements under pressure and performed them with the collaboration of P.G.R. and R.K.K.   Z. B. and J. K. grew and characterized the CaFe$_2$As$_2$ single crystals. R.S.G., M.T., D.D. and G.A.U. developed the 3D BTK model for fitting the experimental PCARS data and made all the data analysis. G.P. made \emph{ab-initio} DFT calculations. K.S. and K.K. made 10-orbital RPA calculations.
R.S.G. and D.D. wrote the paper.

\section*{Additional information}
The authors declare no competing financial interests.

\end{document}